# Graph-based Label Propagation for Semi-Supervised Speaker Identification

*Long Chen, Venkatesh Ravichandran, Andreas Stolcke*

Amazon Alexa, USA

`{longchn, veravic, stolcke}@amazon.com`

## Abstract

Speaker identification in the household scenario (e.g., for smart speakers) is typically based on only a few enrollment utterances but a much larger set of unlabeled data, suggesting semi-supervised learning to improve speaker profiles. We propose a graph-based semi-supervised learning approach for speaker identification in the household scenario, to leverage the unlabeled speech samples. In contrast to most of the works in speaker recognition that focus on speaker-discriminative embeddings, this work focuses on speaker label inference (scoring). Given a pre-trained embedding extractor, graph-based learning allows us to integrate information about both labeled and unlabeled utterances. Considering each utterance as a graph node, we represent pairwise utterance similarity scores as edge weights. Graphs are constructed per household, and speaker identities are propagated to unlabeled nodes to optimize a global consistency criterion. We show in experiments on the VoxCeleb dataset that this approach makes effective use of unlabeled data and improves speaker identification accuracy compared to two state-of-the-art scoring methods as well as their semi-supervised variants based on pseudo-labels.

## 1. Introduction

Deep learning [1] has been shown highly effective across a range of speech processing tasks, including automatic speech recognition [2], speaker recognition and diarization [3], and emotion recognition [4]. However, typical supervised deep learning requires large amounts of training data (as well as corresponding computing resources). It requires large-scale, costly and time-consuming data annotation that is prone to consistency and quality problems. Labeling the identities of unfamiliar speakers from audio data alone is one such challenging annotation task, and presents a major problem for the development of accurate speaker recognition systems.

Semi-supervised learning (SSL) is a technique to reduce the dependency on annotations by learning from unlabeled, as well as labeled, data. SSL has been successfully applied to a variety of fields in machine learning, such as computer vison and natural language processing. And there has been a long-lasting history of innovation in SSL techniques, including pseudo-labeling [5], self-ensembling [6], and virtual adversarial training (VAT) [7].

Recently, graph-based SSL (graph-SSL) methods have received much attention due to their convexity, scalability and unique suitability for capturing intrinsic relationships among datapoints [8]. In graph-SSL, samples (both labeled and unlabeled) are represented as nodes in a weighted graph with edges measuring the similarity between samples. To predict the labels on unlabeled samples by aggregating labels and similarity information throughout the graph, various graph-SSL methods have been developed, such as label propagation (LP) [9]–[11], modified adsorption method [12], and graph convolutional networks (GCN) [13]. Among them, label propagation, one of the simplest kinds of graph-SSL methods, works by propagating label information from labeled to unlabeled nodes over the graph based on sample similarity weights. LP methods typically conduct the propagation in an iterative manner, converge quickly and have lower cost than other deep learning methods. Successful applications to various tasks in computer vision [14], [15] and natural language processing (NLP) [16] have been reported. More recently, Huang et al. [17] demonstrated that graph-SSL methods based on LP can exceed or nearly match the performance of state-of-the-art graph neural networks (GNNs) [13], [18], [19] on a wide variety of benchmarks, with much less parameters and runtime.

In the field of speaker recognition, tasks are usually classified into two categories: speaker verification (SV) and speaker identification (SID). SV verifies whether a given utterance matches a speaker based on the known utterances from the speaker. In SV tasks, embeddings are generated for test utterances as well as for reference utterances and a similarity score, such as cosine distance, is employed to produce a discriminant score. SID means identifying the speaker of each utterance from a fixed set of known speakers. In most cases, SID can be regarded as an *N*-way classification problem, where *N* represents the number of speakers. In much of the research literature, SID models are trained as speaker classifiers on the full set of known speakers, typically employing a fully connected classifier and requiring pre-defined classes. However, in the case of AI smart speakers (e.g., Amazon Echo and Google Home), the devices are typically used by multiple speakers within a household. Thus, the SID task for the household use case involves a large number of disjoint speaker sets, each with a small number of classes, which is similar to few-shot classification [20].

In this work, we propose a graph-SSL method based on label propagation for speaker identification, inferring labels by leveraging unlabeled data. Wang et al. [21] have proposed similar adaption of graph-SSL for speaker diarization, on data from meetings. To the best of our knowledge, our work is the first attempt to apply graph-based SSL for speaker identification in a household scenario. In contrast to other recently proposed supervised [22]–[27] or semi-supervised [28], [29] approaches in speaker recognition that focus on generating better embeddings by leveraging advanced network architectures or loss functions, data augmentation or adversarial training, our approach focuses on speaker label inference (scoring) given an existing speaker embedding extractor and provides a simple, low-cost solution to improve label prediction without tuning the embeddings. Moreover, unlike the mentioned methods [22]–[29], which predict labels individually without considering the similarities among all data samples, our method considers pair-wise scores for all samples in making a prediction, thereby improving SID accuracy.

## 2. Related Work

### 2.1. Semi-Supervised Learning (SSL)

Pseudo-labeling is a simple but powerful implementation of SSL by Lee et al. [5] that outperformed conventional methods on the MNIST test dataset by employing entropy regularization [30]. Self-ensembling [6] methods have also improved the state of the art by using consensus prediction of unknown labels using drop-outs and temporal learning across epochs. In 2018, Miyato et al. [7] proposed a new regularization-based method named virtual adversarial training (VAT) that ensured local smoothness of the conditional label distribution given input perturbations. These methods demonstrate the power of SSL on many popular deep learning tasks.

In a recent survey paper [31], deep learning on graphs has been described as a fast-developing research field. A few graph-based SSL methods have been suggested in the field of speech processing. Yuzong et al. [32] demonstrated the power of graph-based SSL systems by improving phone and segment classification by 3.64% (absolute) over their baseline classifier using their "prior based" measure propagation method on the TIMIT database. Similarly, graph-based learning (GBL) algorithms [33], [34] have been shown to improve the state of the art over supervised algorithms in phonetic classification.

### 2.2. Speaker recognition

Most research in speaker recognition focuses on training a better embedding extractor to encode the speakers' utterances. Recently, advanced network architectures have been investigated for improving speaker embeddings. For example, VGG-M [35], VGGVox [24], AutoSpeech [25], Magneto [22] all utilize CNN-based backbone networks to learn speaker embeddings from pre-processed spectrograms of utterances. GE2E [23] and its variant with attention (GE2E-Att) [27] utilize RNN-based backbone networks to learn speaker embeddings through metric learning. Self-attentive adversarial speaker-identification (SAASI) [26] utilizes self-attention to learn robust embeddings with adversarial training. SSL methods have also been investigated for speaker recognition. Generalized contrastive loss (GCL) [28] combines supervised metric learning and unsupervised contrastive learning with augmentation. Cosine-distance virtual adversarial training (CD-VAT) [29] utilizes VAT to ensure the robustness of the embedding against input perturbations, as measured by cosine distance. Graph-SSL for speaker diarization has shown promising results on speaker attribution [21], for meeting recordings. In contrast to [21], we focus on the SID task and SSL in particular, by testing different embeddings, controlling the amount of unlabeled/labeled data, and comparing it against commonly used baseline SSL/non-SSL methodologies.

## 3. Methods

### 3.1. Problem setup

Let us assume a household with $C$ speakers (classes). Let $(x_1, y_1) \dots (x_l, y_l)$ be the labeled utterances, where $Y_L = \{y_1 \dots y_l\} \subset \{1 \dots C\}$ are the speaker labels. Let $(x_{l+1}, y_{l+1}) \dots (x_{l+u}, y_{l+u})$ be the unlabeled utterances, where $Y_U = \{y_{l+1} \dots y_{l+u}\} \subset \{1 \dots C\}$ are the unknown speaker labels. Let $X = \{x_1 \dots x_{l+u}\} \subset R^D$ be the embeddings of the utterances. The problem is to predict $Y_U$ from $X$ and $Y_L$.

### 3.2. Graph construction

We create a fully connected graph for each household where each node represents an utterance and each edge connecting two nodes as a weight quantifies the similarity between two utterances by its edge weight. The number of nodes in a graph equals the number of (labeled or unlabeled) utterances in the household. There are various ways to measure the similarity between two utterances. Here we use Euclidean distance between the embeddings of the utterances to define the edge weight between two nodes $i, j$:

$$W_{ij} = \exp\left(-\frac{\sum_{d=1}^{D}(x_i^d - x_j^d)^2}{\sigma^2}\right) \quad (1)$$

where $\sigma$ is a temperature-like hyperparameter of the model and $W$ is the matrix of edge weights.

### 3.3. Label propagation

Label propagation (LP) [9]–[11] is a transductive learning approach by which the known labels are propagated to the unlabeled nodes. The basic idea is that given a graph and a small number of nodes with known labels, we want to find a joint labeling of all nodes in the graph such that 1) the labeling is smooth over the graph and 2) the labels that are given a priori are not changed, or by too much. This is typically achieved by minimizing a loss function with two factors: a) supervised loss over the labeled instances, and b) a graph-based regularization term to ensure that the predictions for similar nodes are similar. Here we employ the following objective function:

$$\operatorname{argmin}_f \|\mathbf{f} - \mathbf{Y}\|_2^2 + \lambda \mathbf{f}^T L_{sym} \mathbf{f} \quad (2)$$

where $Y$ is the input vector of known labels, $f$ is the labeling solution, and $\lambda$ is a regularization hyperparameter of this model. $L_{sym}$ is the symmetric normalized Laplacian matrix of the graph: $L_{sym} = I - D^{-1/2}WD^{-1/2}$, where $D$ is the degree diagonal matrix with $D_{ii} = \sum_{j=1}^{l+u} W_{ij}$. The first term of the objective function is the supervised loss and the second term is the graph-regularization term that ensures smoothness, i.e., label consistence of nearby samples. To solve this objective function for each household, we employ the iterative algorithm as introduced by Zhou et al. [10]. This method aims to spread every sample's label information through the graph until achieving global convergence. Compared to the original algorithm, we add a class normalization operation which applies to labels and pseudo-labels in the LP process in order to minimize the influence of imbalance in the labels/pseudo-labels [36]. Algorithm 1 summarizes the label propagation process.

| Algorithm 1: Label Propagation with Normalization |
|---|
| Compute the affinity matrix $W$ as eq. 1 if $i \neq j$ & $W_{ii} = 0$ |
| Compute matrix $S = D^{-1/2}WD^{-1/2}$ |
| Initialize $\hat{Y}^{(0)}: \hat{Y}_{ij}^{(0)} = \begin{cases} 1 & (i \leq l, \ x_i \text{ is labeled as } j) \\ 0 & (else) \end{cases}$ |
| Normalize $\hat{Y}^{(0)}: \hat{Y}_{ij}^{(0)} = \hat{Y}_{ij}^{(0)} / \sum_k \hat{Y}_{kj}^{(0)}$ |
| Choose a parameter $\alpha \in (0,1)$ |
| Iterate $\hat{Y}^{(t+1)} = \alpha S \hat{Y}^{(t)} + (1-\alpha)\hat{Y}^{(0)}$ until convergence |
| Label each point $x_i$ by $y_i = argmax_{j \leq C} \hat{Y}_{ij}^{(\infty)}$ |

# 4. Experiments

## 4.1. Datasets

We used the VoxCeleb2 [24] dataset to train the speaker embedding generator and VoxCeleb1 [35] to construct graphs and evaluate speaker identification performance with different LP methods. Table 1 shows the statistics of the datasets.

Table 1: *Statistics of the datasets.*

| Dataset | VoxCeleb1 | VoxCeleb2 |
|---|---|---|
| # of speakers | 1,251 | 6,112 |
| # of male speakers | 690 | 3,761 |
| # of utterances | 153,516 | 1,128,246 |
| Avg# of utterances per speaker | 116 | 185 |
| Avg length of utterances (s) | 8.2 | 7.8 |

## 4.2. Experimental setup

For embedding generator training, we trained our models in the text-independent speaker verification scenario as introduced in the GE2E [23] paper. We also trained another embedding generator with the GE2E-Att architecture [27], a variant of GE2E with an attention layer on top of an LSTM to produce more informative embeddings.

For evaluating model performance on a speaker identification task, the experiments are conducted in a simulated household scenario, simulating the use case of most smart speaker AI assistants. The 1251 speakers in VoxCeleb1 are randomly shuffled and sampled without replacement into 312 households with each household comprising 4 speakers. We further split the 312 households into 112 households as the development set and the remaining 200 as the validation set. The development set is used for optimizing hyperparameters for our approach and the validation set is used for final evaluation. After hyperparameter optimization we set $\sigma = 0.22$ in Equation 1 and $\alpha = 0.99$ in Algorithm 1.

For each household, 10 utterances per speaker are randomly selected to serve as the holdout dataset for evaluation. The rest of the utterances can be selected either as labeled samples (aka enrollment utterances) or unlabeled samples for the SSL experiments. We use the speaker identification error rate (SIER) within a household as the metric to evaluate performance. SIER is defined as 1 – (accuracy of top predicted speaker). The final SIER is calculated as the micro-average over the 200 households in the validation set.

## 4.3. Methods comparison

The main focus of this study is to investigate the proposed label propagation algorithm for an accurate speaker classification in the presence of prior unlabeled samples. In real-time, each household contains a different set of speakers. Thus, conventional speaker identification approaches which require the pre-defined classes or are trained with a fully connected classifier, are not suitable. In practice, speaker identification in a household scenario is usually treated as an SV task, where the predicted label for each utterance is given by the household speaker with the highest speaker verification (SV) score.

We evaluate LP-based approaches against four baselines:

- **CS**: Cosine scoring [37], [38]. We compute the cosine similarity score between each utterance in the holdout dataset and each labeled utterance for each speaker, and compute an average score per speaker. The speaker with the highest score is picked as the predicted label for the holdout utterance. No prior unlabeled utterances are used.
- **CSEA**: Cosine scoring with embedding average [23], [39]. For each speaker/class, we compute the speaker level representation by averaging across all embeddings from the labeled utterances belong to the speaker. For each utterance in the holdout dataset, we compute the cosine similarity score to the speaker-level representation. The speaker with the highest score is picked as the predicted label for the utterance in the holdout dataset. No prior unlabeled utterances are used.
- **2-CS**: A 2-step cosine scoring. In Step 1, we calculate pseudo-labels using the CS method for all unlabeled utterances. In Step 2, we predict the labels of utterances in the holdout dataset with the labeled utterances and pseudo-labels from Step 1, using the CS method.
- **2-CSEA**: A 2-step cosine scoring with embedding average. In Step 1, we calculate pseudo-labels using the CSEA method for all unlabeled utterances. In Step 2, we predict the label for the holdout dataset with labeled utterances and pseudo-labels from Step 1, using CSEA.

We evaluated the following three LP-based methods:

- **LP**: Simple label propagation over an undirected graph containing labeled, unlabeled and holdout samples for each household (Algorithm 1).
- **2-LP**: A 2-step label propagation: In this method, Step 1 performs label propagation to compute predictions for each unlabeled datapoint using labeled datapoints. In Step 2, we utilize the predictions from Step 1 as pseudo-labels and perform a second round of label propagation to make final predictions for each utterance in the holdout dataset, using labeled and unlabeled data.
- **2-LPEA**: A 2-step method with label propagation as Step 1 and embedding average as Step 2. Step 1 performs label propagation to compute predictions for each unlabeled datapoint using labeled data. In Step 2, we utilize the predictions from Step 1 as pseudo-labels and use the CSEA method to make final predictions for each utterance in the holdout dataset, using labeled and unlabeled datapoints.

Rationales for comparing these methods are: 1) CS and CSEA are the most commonly used methods for speaker verification in previous works, but they do not use unlabeled data for prediction. 2) 2-CS and 2-CSEA extend CS and CSEA by using traditional pseudo-labels on unlabeled data, giving us a suitable baseline for other SSL methods. 3) 2-LP extends LP by converting soft labels to hard labels for unlabeled samples and then normalizing the labels and pseudo-labels in the second step to have more a balanced label distribution for the classification of holdout data. 4) 2-LPEA is a pragmatic approach with fixed runtime computation. Step 1 followed by embedding averaging computes a compact speaker representation offline, while benefitting from LP. Runtime processing involves only traditional scoring of embeddings.

## 4.4. Experimental results

For evaluation, we conducted experiments by randomly selecting $L$ utterances per speaker as the labeled samples and $U$ utterances per household as the unlabeled samples.

Table 2: *SIER (%) on validation set with GE2E and GE2E-Att embeddings (L=2).*

| Method | GE2E | | GE2E-Att | |
|---|---|---|---|---|
| | $U$=40 | $U$=All | $U$=40 | $U$=All |
| CS | 3.36 | 3.36 | 2.28 | 2.28 |
| CSEA | 3.06 | 3.06 | 2.08 | 2.08 |
| 2-CS | 2.05 | 1.69 | 1.18 | 1.01 |
| 2-CSEA | 1.93 | 1.39 | 1.15 | 0.87 |
| LP | 1.82 | 1.38 | 1.00 | 0.77 |
| 2-LP | 1.73 | **1.25** | 0.94 | **0.69** |
| 2-LPEA | **1.49** | 1.31 | **0.88** | 0.84 |

Table 2 summarizes the speaker identification error rates (SIER) for different label prediction methods on the validation set. Here, for each household, $L$=2 and $U$=40 or "All", where "All" refers to the case that all remaining utterances except the holdouts were selected as the unlabeled samples. To verify whether the proposed approaches are generic to different embeddings, we conducted experiments on two sets of 512-dimentional embeddings trained with GE2E [20] and GE2E-Att [24] methods, respectively. As shown in Table 2, the proposed methods with label propagation outperform all the baseline methods for both GE2E and GE2E-Att embeddings. The 2-LP method achieves the lowest SIER when there are enough unlabeled samples ($U$=All) with an improvement of 10.1% and 20.7% against the best baseline methods for GE2E and GE2E-Att embeddings, respectively. The 2-LPEA achieves the lowest SIER where there is a relatively small amount of unlabeled data ($U$=40) with an improvement of 22.8% and 23.5% against the best baseline methods for GE2E and GE2E-Att embeddings, respectively.

Figure 1 shows how SIER varies with different maximum number of unlabeled samples per household (on the validation set with the GE2E embeddings). Here, for each household, $L$=2 and $U$ = 0, 40, 80, 160, 320, 640 or "All". We observe that all methods utilizing unlabeled data outperform those that do not (CS and CSEA). With few exceptions results improve with added unlabeled data. This result demonstrates the power of using unlabeled data and SSL for speaker prediction, even with simple pseudo-labeling. Moreover, LP-based methods have the lowest SIER with different amounts of unlabeled data, even for the case $U$=0. This fact demonstrates that LP is very effective at improving SIER by utilizing graph-regularization, even when applied only to holdout samples (the case of $U$=0).

Figure 2 shows the SIER with varying numbers of labeled samples ($L$=1, 2, 4, 10, 20, 40, All) per speaker (on the validation set with the GE2E embeddings), where all remaining utterances are used as the unlabeled samples. As can be expected, all methods achieve better performance with more labeled samples. LP-based methods have the lowest SIER regardless of the amount of labeled data. As the number of labeled samples increases, the performance gain with the 2-step methods decreases to a negligible value (2-step methods merge with the 1-step method, e.g., 2-CSEA merges with CSEA). However, methods with LP as the final step (LP and 2-LP) still outperform the other methods by substantial margin, even for the case with sufficient amount of labeled data ($L$=All).

Comparing the three LP-based methods, 2-LP performs better than LP, which indicates the benefits of using hard labels and normalization in the second step. 2-LPEA works the best for relatively small numbers of labeled and unlabeled samples.

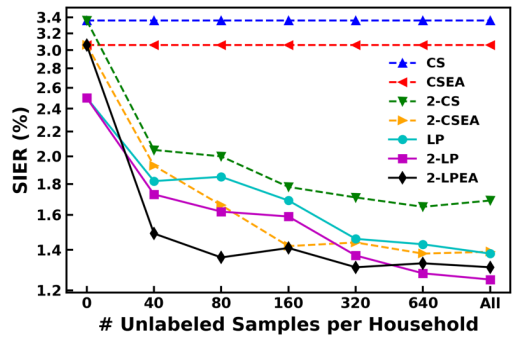

Figure 1: *SIER (%) in log scale as a function of the number of unlabeled samples per household*

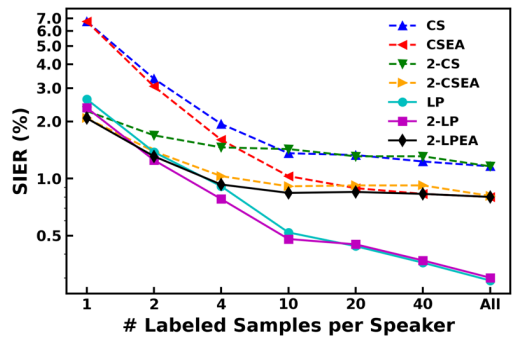

Figure 2: *SIER (%) in log scale as a function of the number of labeled samples per speaker*

That is because embedding averaging is intrinsically more robust to the defects in the pseudo-labels. However, for the case with sufficient numbers of unlabeled or labeled samples, when high-quality pseudo-labels can be derived from in Step 1, 2-LP outperforms all other methods.

## 5. Conclusions

We have proposed a generic semi-supervised learning approach to improve speaker identification accuracy by label propagation on a graph encoding pairwise similarity for all labeled and unlabeled utterances. We evaluated several variants of the proposed method on VoxCeleb datasets in simulated household scenarios with varying amounts of labeled/unlabeled data and different embedding methods. Evaluation against the baseline methods demonstrate that the proposed approaches are very effective in lowering SID error under various conditions.

Unlike other supervised or semi-supervised methods which typically improve the embeddings, this approach provides a simple and low-cost solution to make better predictions without tuning the embeddings, utilizing all speaker similarity scores available within a household. We conclude that graph-based label propagation is a versatile and effective method to improve speaker identification.

## 6. Acknowledgments

We would like to thank Charles Chang for supporting this work. We would also like to thank the scientists and engineers at Amazon Alexa Voice Recognition group (Speaker ID team) for valuable input.